\documentclass[prx,aps,twocolumn,showpacs,superscriptaddress,amsmath,amssymb]{revtex4}

\usepackage{graphicx}
\usepackage{psfrag}
\usepackage{epsfig}
\usepackage{color}
\usepackage{subfigure}
\definecolor{dred}{rgb}{0.7,0.0,0.0}
\usepackage{dcolumn}% Align table columns on decimal point
\usepackage{bm}% bold math

\begin{document}

%
% Title Page
%

%\title{Isotropic Quenched Disorder Enhances Nematic Tendencies\\
%in Electron-Doped Pnictides.}
  
%\title{Quenched disorder triggers a robust nematic state \\
%in electron-doped 122 pnictides.}
%
\title{ Isotropic Quenched Disorder Triggers a Robust Nematic State \\
in Electron-Doped Pnictides.}

\author{Shuhua Liang} 
\author{Christopher B. Bishop}
\author{Adriana Moreo}
\author{Elbio Dagotto}
\affiliation{Department of Physics and Astronomy,University of Tennessee,
Knoxville, TN 37966, USA} 
\affiliation{Materials Science and Technology Division,
Oak Ridge National Laboratory,Oak Ridge, TN 37831, USA}

\date{\today}
%\maketitle

\begin{abstract}
{The phase diagram of electron-doped pnictides is studied varying the
temperature, electronic density, and isotropic quenched disorder strength by means 
of computational techniques applied to a three-orbital ($xz$, $yz$, $xy$) spin-fermion model with lattice degrees 
of freedom. In experiments, chemical doping introduces disorder but in theoretical studies
the relationship between electronic doping and the randomly located dopants, with their associated quenched
disorder, is difficult to address. 
In this publication, the use of computational techniques allows us to study independently the effects
of electronic doping, regulated by a global chemical potential, and impurity 
disorder at randomly selected sites.
Surprisingly, our Monte Carlo simulations reveal that the fast reduction with doping
of the N\'eel $T_N$ and the structural $T_S$ transition temperatures, and 
the concomitant stabilization of
a robust nematic state, is primarily
controlled by the magnetic dilution associated with the in-plane isotropic disorder
introduced by Fe substitution. In the doping range studied, 
changes in the Fermi Surface produced by electron doping 
affect only slightly both critical temperatures.
%
%Actually, the separation between
%$T_N$ and $T_S$, leading to the stabilization of a robust nematic phase in the doped system, 
%is triggered by caused by disorder, otherwise $T_N$ and $T_S$ would remain be the same
%in the range of doping investigated.
%In addition, it is observed that anisotropic magnetic clusters form upon the introduction of isotropically symmetric impurities.
%The region above $T_S$ is characterized by a coexistence of finite domains
%with magnetic order $(\pi,0)$ and $(0,\pi)$. The domains are elongated along their antiferromagnetic (AF) direction. The magnetic 
%domains coexist with paramagnetic clusters that form around the impurities. As a result, the magnetic structure factor, S({\bf k}), develops equal
%weight peaks at momentum $(\pi,0)$ and $(0,\pi)$, but the peaks are broader (narrower) along the
%corresponding ferromagnetic (antiferromagnetic) direction which correlates with the amplitude of the fluctuations. 
%At $T_S$ the length of the domains is about 8 lattice sites along the AFM direction. 
%%and a small incommensurability is detected with peaks observed at $(\pi\pm\delta_x,\pm\delta_y)$ with $\delta_y>\delta_x$. 
%As $T$ decreases the size of the short range AFM clusters that characterize the nematic 
%phase increases and 
%at the N\'eel temperature, as expected, a divergent peak develops at $(\pi,0)$ in S({\bf k}). 
Results obtained varying the disorder strength indicate 
that the specific material dependent phase diagrams experimentally observed
are a consequence of the variation in disorder profiles introduced by the different 
dopants. Our results are also in agreement 
with neutron scattering and scanning tunneling microscopy, unveiling a patchy network of locally magnetically
ordered clusters with anisotropic shapes, even though the quenched disorder is locally isotropic.  
%It is found that the electronic and magnetic anisotropies 
%that characterize the nematic phase of electron-doped 122 pnictides are stabilized by isotropic disorder.
This study reveals 
a remarkable and unexpected degree of complexity in pnictides: the fragile tendency to nematicity
intrinsic of translational invariant electronic systems needs to be supplemented by quenched 
disorder to stabilize the robust nematic phase experimentally found in electron-doped 122 compounds.}
\end{abstract}
 

 
\maketitle
 
\section{Introduction}

The mechanism that leads to high critical temperature superconductivity 
in iron-pnictides~\cite{Fe-SC,peter,fernandes.nat.phys,pengcheng,dagottoRMP13}
is still elusive, mainly because the several simultaneously active
degrees of freedom (d.o.f.) in these materials pose a major theoretical challenge. 
While magnetic mechanisms are often invoked to explain the $d$-wave 
superconductivity in the cuprates~\cite{doug,cuprates}, the role of the orbitals is added to the 
mix in the case of the iron-based compounds. Moreover, 
the symmetry of their superconducting state is still under considerable debate~\cite{xcheng}. 

The interaction among the many different d.o.f. in pnictides generates rich phase diagrams 
when varying temperature and doping~\cite{phases}. In addition to the superconducting phase, 
magnetic and nematic phases, accompanied by structural distortions, 
have been identified~\cite{phases,fisher.science1,fisher.science2,kuo,fisher1}. 
To properly address this difficult problem it is necessary that 
the spin, orbital, lattice, and charge should all be incorporated in a treatable model 
that allows to monitor their respective roles in the properties of these materials.  
Due to the complexity of the problem most of the previous 
theoretical studies have been performed either in the weak 
or strong coupling limits. 
In weak coupling, the interactions among the electrons are considered 
small and the physical properties are studied in momentum space in terms of 
itinerant electrons, with emphasis on particular properties of their Fermi Surfaces 
(FS) such as nesting~\cite{spin3,fernandes1,fernandes2,kontani}.
On the other hand, the strong coupling approach is based on the experimental 
observation of localized magnetic moments and on the fact 
that several properties of the pnictides can be reproduced via 
Heisenberg models~\cite{si,spin1,spin2}.  
Both approaches were successful in the study of the magnetic properties 
of the parent compounds, indicating that in these materials both localized and itinerant
magnetic moments are important. However, upon doping there are challenges 
explaining experimental data in both approximations. In particular, 
when doping is achieved by chemical substitution of iron atoms then
the associated effects of disorder must also be incorporated into the theoretical considerations. 

The parent compound of the 122 family, BaFe$_2$As$_2$, can be doped with electrons 
by replacing Fe by a transition metal (TM) resulting in Ba(Fe$_{1-x}$TM$_x$)$_2$As$_2$ or 
with holes by replacing Ba by an alkali metal (A) leading to Ba$_{1-x}$A$_x$Fe$_2$As$_2$~\cite{xcheng}. 
It is also possible to dope in an isovalent manner replacing, for example, Fe with Ru to obtain 
Ba(Fe$_{1-x}$Ru$_x$)$_2$As$_2$~\cite{rullier}. Nominally, replacing Fe with Ru, Co, Ni, and Cu 
would introduce 0, 1, 2, and 3 electrons per dopant atom. However, experiments indicate
a difference between nominal doping $x$ and the measured doping concentration $x_m$ usually 
determined using wavelength dispersive x-ray spectroscopy (WDS)~\cite{phases}. This 
means that in some cases, electrons can get trapped by the doped impurities~\cite{sawatzky}. 
Chemical substitution introduces an amount of disorder that is difficult to control 
experimentally. In addition to electrons being trapped, other effects such as magnetic 
dilution and impurity scattering may also occur~\cite{hirschfeld}. 

In undoped 122 compounds the structural and the N\'eel transition temperatures, $T_S$ and $T_N$, 
are equal to each other. Upon electron doping both are rapidly reduced, 
with $T_S$ decreasing at an equal 
or slower rate than $T_N$~\cite{rullier,phases}. The reduction of 
these temperatures is explained in weak coupling by a loss 
of FS nesting induced by the electronic doping and in strong coupling by magnetic 
dilution as in $t$-$J$ models. 
However, these views seem to be in contradiction with 
several experimental results. For example, in Ba(Fe$_{1-x}$Ru$_x$)$_2$As$_2$, which nominally does 
not introduce electronic doping and associated changes in FS should not be expected, 
both $T_S$ and $T_N$ decrease with doping and the material eventually becomes 
superconducting~\cite{rullier}. In addition, doping with Co, Ni, and Cu is 
expected to introduce 1, 2, and 3 extra electrons per doped atom. However, the 
experimentally observed reduction on $T_N$ and $T_S$ was found to be primarily a function of 
the doping $x$ rather than of the density of electrons~\cite{phases,kuo-fisher}. Experiments, 
thus, indicate that when dopants are introduced directly on the Fe-As planes,
as it is the case for electron-doped 122 materials, disorder must play 
an important role~\cite{nakajima,phases,kontani,chuang,allan}.
Due to the experimental uncertainty on the actual doping concentration 
and the nature of the disorder, a theoretical understanding of the phase diagrams under these challenging 
circumstances is elusive. Density functional theory (DFT) studies indicated that
in-plane-doped atoms would tend to trap electrons~\cite{sawatzky}, while 
first-principles methods found that the interplay between on-site and off-site 
impurity potentials could induce FS distortions in nominally 
isovalent doping~\cite{hirschfeld}. Moreover,  a calculation considering 
two-orbiton processes predicts a non-symmetric impurity potential which could be 
responsible for the observed transport anisotropies~\cite{kontani}.     

In this publication, the effects of electron doping in the 122 pnictides 
will be studied numerically using a spin-fermion model (SFM) 
for the pnictides~\cite{kruger,BNL,shuhua} including the lattice d.o.f.~\cite{shuhua13}.
The SFM considers phenomenologically  the experimentally gathered evidence that requires a
combination of itinerant and localized d.o.f. to properly 
address the iron-based superconductors~\cite{pengcheng,dagottoRMP13,loca1,loca2}. 
The itinerant sector mainly involves
electrons in the $xz$, $yz$, and $xy$ $d$-orbitals~\cite{three} while
the localized spins represent the spin of the other $d$-orbitals~\cite{kruger,BNL}, or
in a Landau-Ginzburg context it can be considered as the magnetic order parameter. 

The focus of this effort will be on the structural and the N\'eel transitions, 
and the properties of the resulting nematic phase that will be monitored 
as a function of the electronic and impurity densities.
Earlier studies performed in the undoped parent compounds indicated that the coupling between 
the lattice orthorhombic distortion $\epsilon_{\bf i}$, associated to 
the elastic constant C$_{66}$, and the spin-nematic order parameter  
$\Psi_{\bf i}$ stabilizes the orthorhombic $(\pi,0)$ antiferromagnetic (AFM) ground state~\cite{shuhua13} 
with $T_S=T_N$ as in the 122 materials~\cite{phases}. The small separation 
between $T_S$ and $T_N$ observed in the parent compounds of the 1111 
family~\cite{clarina} was found to be regulated
by the coupling of the lattice orthorhombic distortion to 
the orbital order parameter $\Phi_{\bf i}$~\cite{shuhua13}. 
%As a consequence, here only the coupling
%between the lattice and the spin-nematic order parameter will be considered because our focus
%are the 122 materials.
 
This is the first time that electronic doping is computationally studied
in a system that includes magnetic, charge, orbital, and lattice d.o.f. supplemented by quenched disorder. 
Our  numerical approach involves Monte Carlo (MC) calculations
on the localized spin and lattice components, combined with a fermionic
diagonalization of the charge/orbital sector. In addition, twisted boundary 
conditions (TBC) and the Travelling Cluster 
Approximation (TCA) are implemented~\cite{chris} in order to study large clusters 
of size $64 \times 64$, a record for the spin-fermion model. 
This numerical approach allows us to incorporate the effects of random on-site and off-diagonal 
disorder and to obtain results for temperatures above $T_S$ where all d.o.f. 
develop strong short-range fluctuations~\cite{spin3,egami}, a regime difficult 
to reach by other many-body procedures. Our main conclusion is that quenched disorder 
is needed to enhance the (weak) electronic tendency to form 
a nematic phase in 122 materials. That a critical temperature such as $T_N$ 
decreases faster with doping by including disorder than in the clean limit is natural~\cite{vavilov1,vavilov2}, but our most novel result is the concomitant stabilization of a nematic regime. In other
words, $T_N$ and $T_S$ are affected {\it differently} by disorder. Isotropic disorder
is sufficient to obtain these results.
Our analysis illustrates
the interdependence of the many degrees 
of freedom present in real materials and the need to study
models with robust many-body techniques to unveil the physics that emerges in these complex systems. 

The organization of the paper is as follows: the model is described in Section~\ref{spin-fermion model} 
and the computational methods are presented in Section~\ref{methods}. 
Section~\ref{results} is devoted to the main results addressing the phase diagram upon doping. 
Section~\ref{nematicproperties} describes the properties of the nematic phase stabilized in our study,
including a comparison with neutron scattering and scanning tunneling microscopy experiments. 
The discussion and summary are the scope of Section~\ref{discussion}. 

\section{Model}\label{spin-fermion model}

\subsection{Hamiltonian}

The spin-fermion model Hamiltonian studied here is based
on the original purely electronic model~\cite{kruger,BNL,shuhua} supplemented by the recent addition of couplings to the lattice degrees of freedom~\cite{chris}:
\begin{equation}
H_{\rm SF} = H_{\rm Hopp} + H_{\rm Hund} + H_{\rm Heis} + H_{\rm SL} + H_{\rm Stiff}.
\label{ham}
\end{equation}
\noindent $H_{\rm Hopp}$ is the three-orbitals ($d_{xz}$, $d_{yz}$, $d_{xy}$) tight-binding Fe-Fe hopping of electrons, with the
hopping amplitudes selected to reproduce ARPES experiments. Readers can find these
amplitudes in previous publications, such as in Eqs.(1-3) and Table 1 
of Ref.~\cite{three}. The average density of electrons per iron and per orbital 
is $n$=4/3 in the undoped limit~\cite{three} and its value in the doped case 
is controlled via a chemical potential
included in $H_{\rm Hopp}$~\cite{chris}. 
The Hund interaction is standard:
$H_{\rm Hund}$=$-{J_{\rm H}}\sum_{{\bf i},\alpha} {{{\bf S}_{\bf i}}\cdot{{\bf s}_{{\bf i},\alpha}}}$,
with ${{\bf S}_{\bf i}}$ the localized spin at site ${\bf i}$ and ${\bf s}_{{\bf i},\alpha}$ 
the itinerant spin corresponding to orbital $\alpha$ at the same site~\cite{foot}.
$H_{\rm Heis}$ contains the Heisenberg interaction among the localized spins involving
both nearest-neighbors (NN) and next-NN (NNN) 
interactions with respective couplings $J_{\rm NN}$
and $J_{\rm NNN}$, and a ratio $J_{\rm NNN}$/$J_{\rm NN}$ = 2/3 (any ratio larger than 1/2
would have been equally effective to favor ``striped'' spin order). Having NN and NNN 
Heisenberg interactions of comparable magnitude arise from having 
comparable NN and NNN hoppings, caused by the geometry of the material.

The coupling between the spin and lattice degrees of freedom is given 
by $H_{\rm SL}$=$-g\sum_{\bf i}\Psi_{\bf i}\epsilon_{\bf i}$~\cite{fernandes1,fernandes2}, 
where $g$ is the spin-lattice coupling~\cite{only}.
Finally, $H_{\rm Stiff}$ is the spin stiffness given by a Lennard-Jones potential 
that speeds up convergence, as previously discussed~\cite{chris}.
Note that the lattice-orbital 
coupling term, $H_{\rm OL}$=$-\lambda\sum_{\bf i}\Phi_{\bf i}\epsilon_{\bf i}$~\cite{chris}, 
is omitted because previous
work indicated that $\lambda$ induces a (small)
nematic phase with $T_S>T_N$ directly in the 
parent compounds~\cite{shuhua13,chris}. Since the goal of the present 
effort is to study the 122 family, 
characterized by $T_S=T_N$ in the undoped case, then this term is not included to reduce the
number of parameters. 

\begin{figure}[thbp]
\begin{center}
\includegraphics[trim = 15mm 0mm 20mm 0mm,width=\linewidth,clip,angle=0]{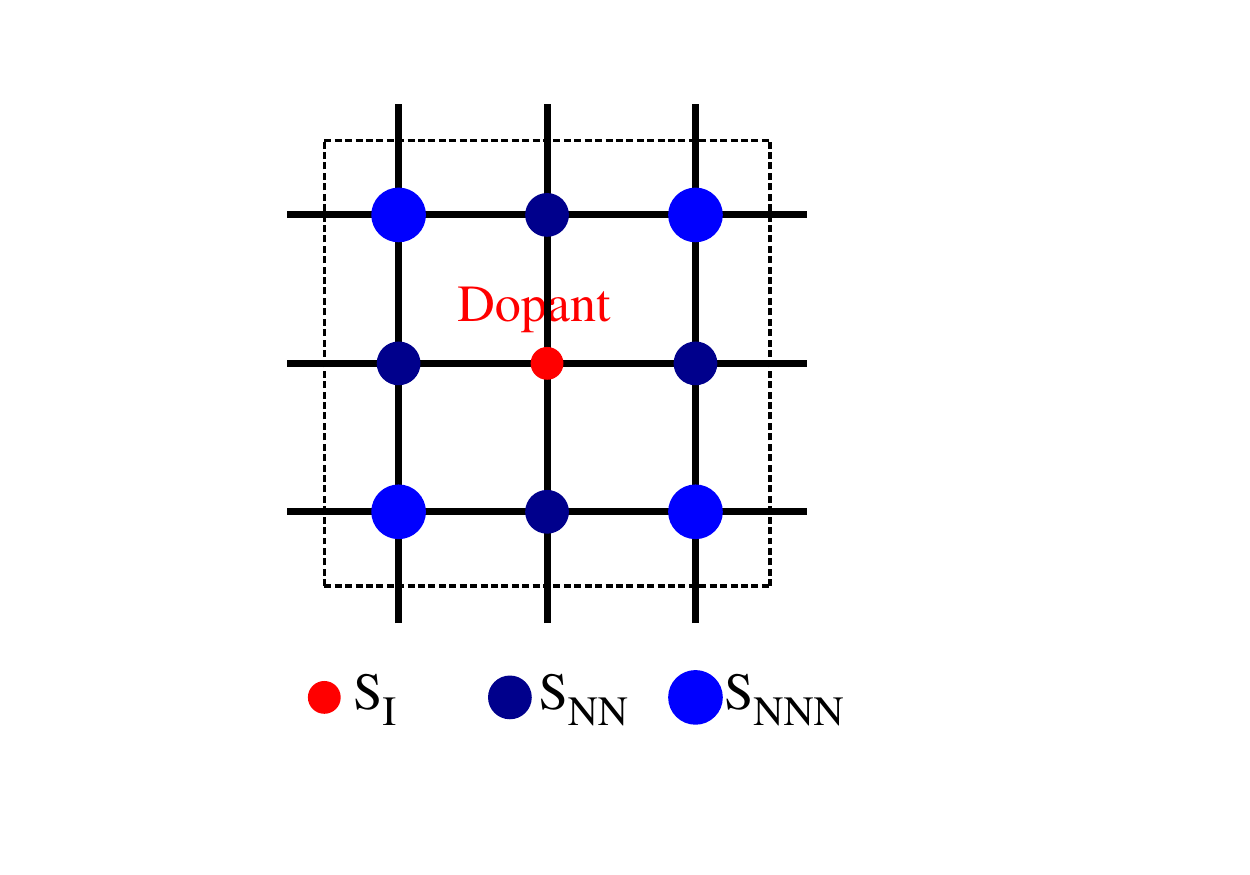}
\vskip -0.3cm
\caption{(color online) {\it Internal structure of dopant sites.} 
Sketch shows the location of a dopant where the magnitude 
of the localized spin, $S_{\rm I}$, is reduced from the original value $S$.
In addition, the neighboring localized spins are also assumed to be
affected by the presence of the dopant. The four immediate nearest-neighbors
have a new localized spin magnitude $S_{\rm NN}$, while the four next nearest-neighbors
have a new localized spin magnitude $S_{\rm NNN}$, such
that $S_{\rm I} \leq S_{\rm NN} \leq S_{\rm NNN} \leq S$ ($S$ is the undoped localized 
spin magnitude, assumed to be 1 in this publication unless otherwise stated).}
\label{vacancyprofile}
\end{center}
\end{figure}

\subsection{Quenched Disorder}

On-site diagonal disorder is introduced by adding an impurity potential $I_I({\bf i}_d)$ to $N_I$ randomly selected sites 
${\bf i}_d$ where transition metal atoms replace Fe. The density of impurity atoms $x$ is 
defined as $x=N_{\rm I}/N$, where $N$ is the total number of lattice sites. 
In addition, the value of the localized spin at the impurity site, $S_I$, is 
reduced since, for example, Co dopants in BaFe$_2$As$_2$ are non-magnetic~\cite{athena}. 
This effectively reduces the local Hund coupling ${J_{\rm H,I}}$ 
and the spin-lattice coupling $g_{\rm I}$ at the impurity sites. 
We also will study the effect of extending the spatial range of the impurity by
reducing the values of the localized spins to $S_{\rm NN}$ ($S_{\rm NNN}$) 
at the NN (NNN) of the impurity 
sites with the corresponding effective decrease in $J_{\rm H}$ and $g$
at those sites (see Fig.~\ref{vacancyprofile}). 
Thus, off-diagonal isotropic disorder results from the effective reduction of 
the Heisenberg couplings at the bonds connecting the impurity
sites and their neighbors~\cite{foot}. Note that off-diagonal disorder could also 
be introduced in the eight hopping amplitudes present 
in $H_{\rm Hopp}$~\cite{chris} but for simplicity we decided not 
to consider hopping disorder at this time.

\section{Methods}\label{methods}

The Hamiltonian in Eq.(\ref{ham}) was studied via a well-known Monte Carlo method~\cite{shuhua,CMR} applied to {\it (i)} 
the localized (assumed classical) spin 
degrees of freedom ${\bf S_i}$ and {\it (ii)} 
the atomic displacements that determine the local 
orthorhombic lattice distortion $\epsilon_{\bf i}$~\cite{shuhua13,chris}. 
For each Monte Carlo configuration of spins and atomic positions the 
remaining quantum fermionic Hamiltonian is diagonalized. The simulations are 
performed at various temperatures, dopings, and disorder configurations and local and long-range observables are measured.
Note that with the exact diagonalization technique 
results can be obtained comfortably only on up to $8\times 8$ lattices, 
which may be too small to provide meaningful data at the low rates of doping 
relevant in the pnictides. For this reason
we have also used the Traveling Cluster Approximation~\cite{kumar} where a larger lattice ($64\times 64$ sites in most of this effort) 
can be studied by performing the MC updates via a 
travelling cluster centered at consecutive sites ${\bf i}$, 
with a size substantially smaller than the full lattice size of the entire system. 
Twisted boundary conditions were also used~\cite{salafranca} 
to obtain (almost) a continuum range of momenta. 
For simplicity, most couplings are fixed to values used
successfully before~\cite{shuhua}: ${J_{\rm H}}$=$0.1$~eV, 
${J_{\rm NN}}$=$0.012$~eV, and ${J_{\rm NNN}}$=$0.008$~eV. 
The dimensionless version of the spin-lattice coupling $\tilde g$  is fixed to 0.16 as in~\cite{shuhua13}. 
The focus of the publication is on the values for the parameters associated 
with disorder and the corresponding physical results, as discussed in the sections below.

An important technical detail is that to improve numerical convergence, and to better mimic real materials that often display an
easy-axis direction for spin orientation, we have introduced a small anisotropy 
in the $x$ component of the super-exchange interaction so that the actual Heisenberg interaction is:
\begin{equation}
\begin{split}
H_{\rm Heis}= J_{{\rm NN}}\sum_{\langle{\bf ij}\rangle} ({\bf S}_{\bf i}\cdot{\bf S}_{\bf j}+\delta S_{\bf i}^xS_{\bf j}^x)\\
+J_{\rm NNN}\sum_{\langle\langle{\bf im}\rangle\rangle} ({\bf S}_{\bf i}\cdot{\bf S}_{\bf m}+\delta S_{\bf i}^xS_{\bf m}^x), 
\label{heis}
\end{split}
\end{equation}
%\vspace{-0.2cm}
\noindent with $\delta=0.1$. This anisotropy slightly raises $T_N$, but the magnetic 
susceptibility $\chi_S$ becomes much sharper at the transition temperatures, facilitating an accurate 
determination of $T_N$.

The Monte Carlo simulations with the TCA procedure were mainly performed using $64 \times 64$ square lattices~\cite{finsize}.
Typically 5,000 MC steps were devoted to thermalization and 10,000 to 25,000 steps for 
measurements at each temperature, doping, and disorder configuration. The results presented below arise from averages over 
five different disorder configurations. The expectation values of  
observables remain stable upon the addition of extra configurations due to self-averaging.  
The magnetic transition was determined by the behavior of the magnetic susceptibility defined as
\begin{equation}
\chi_{S(\pi,0)}=N\beta\langle S(\pi,0)-\langle S(\pi,0)\rangle\rangle^2,
\label{Xs}
\end{equation}
\noindent where $\beta=1/k_BT$, $N$ is the number of lattice sites, and $S(\pi,0)$ is the magnetic structure 
factor at wavevector $(\pi,0)$ obtained via the Fourier transform of the real-space spin-spin 
correlations measured in the MC simulations. The structural transition is determined by the behavior 
of the lattice susceptibility defined by
\begin{equation}
\chi_{\delta}=N\beta\langle \delta-\langle\delta\rangle\rangle^2,
\label{Xa}
\end{equation}
\noindent where $\delta={(a_x-a_y)\over{(a_x+a_y)}}$, and $a_i$ is the lattice constant along the $i=x$ or $y$ directions. 
These lattice constants are determined from the orthorhombic displacements 
$\epsilon_{\bf i}$~\cite{chris}.

\section{Results}\label{results}

Our first task is to understand the effect of doping and disorder on the magnetic
and structural transitions. For this purpose, we studied the evolution of $T_N$ and $T_S$ vs. doping concentration
under different disorder setups.

\subsection{Clean limit}

\begin{figure}[thbp]
\begin{center}
\includegraphics[trim = 10mm 0mm 10mm 0mm,width=\linewidth,clip,angle=0]{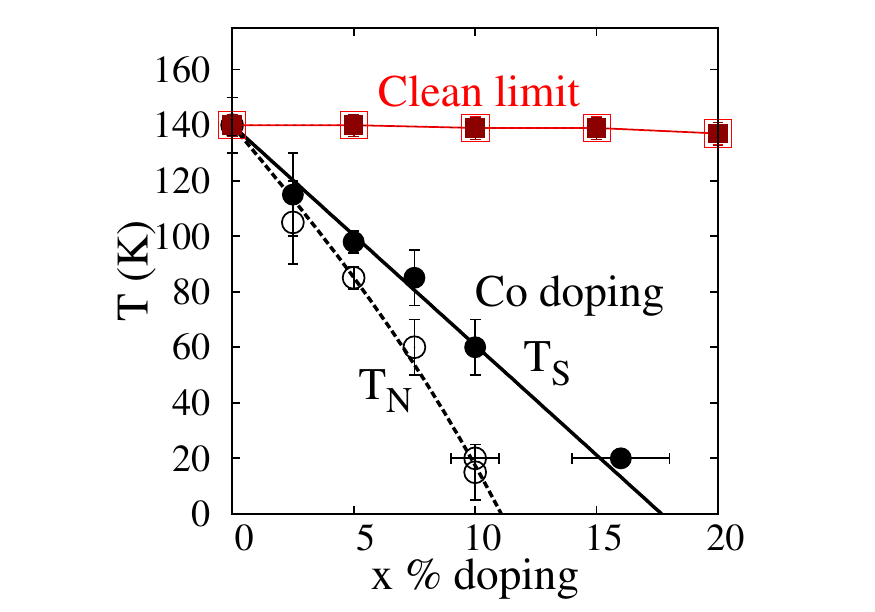}
\vskip -0.3cm
\caption{(color online) {\it Clean limit and effect of Co doping.} 
The clean limit results (open and solid red points) indicate that
$T_S=T_N$ and both are approximately constant in the range studied. 
For Co doping, the N\'eel temperature $T_N$ (open circles and black dashed line) 
and the structural transition temperature $T_S$ 
(filled circles and black solid line) vs. the percentage of impurities $x$ are shown. 
The on-site disorder is $I_{\rm I}=-0.1$ and the off-diagonal disorder is 
determined by $S_{\rm I}=0$, $S_{\rm NN}=S/4$, and $S_{\rm NNN}=S/2$. 
For both sets of curves, i.e. with and without quenched disorder, 
the density of doped electrons equals $x$ to simulate Co doping. 
The cluster used has a size $64\times64$.}
\label{xTJ}
\end{center}
\end{figure}

Consider first the ``clean limit''. The red squares in Fig.~\ref{xTJ} show the 
evolution of $T_N$ and $T_S$ when
the electronic doping does $not$ introduce disorder. 
In this case $T_N$ is hardly affected and it continues to be equal to $T_S$ for all dopings investigated here. 
This result indicates that the reduction of $T_N$ and $T_S$, 
and the stabilization of a nematic phase 
in between the two transitions observed experimentally 
upon electron doping~\cite{phases}, does not emerge 
just from the reduction of Fermi Surface nesting 
induced by the electronic doping. This conclusion is not surprising 
if we recall that the undoped $N$-site lattice 
has $4N$ electrons which means that for $x= 10\%$ 
the number of added electrons is $N_e=0.1N$ and, thus, 
the percentual change in the electronic density 
is just $100 \times (0.1N/4N)=2.5$\%. Such a small percentual 
variation in the electronic density should 
not produce substantial modifications in the FS,  
explaining why the changes in nesting 
are small and, thus, why the critical temperatures are not 
significantly affected. Then, disorder maybe needed to understand the experiments.

\subsection{Co doping}

To study the effect of quenched disorder, let us first focus on Co doping, 
which nominally introduces one extra electron per dopant. 
In Fig.~\ref{xTJ}, the N\'eel and structural transition temperatures 
are presented for the case where one extra electron is contributed by each replaced iron atom, 
which means that $x=n$, where $n$ is the density of added electrons
and $x$ is the density of replaced iron atoms. We considered several possible values for the 
on-site impurity potential and spin values near the impurity (see details 
discussed below) and we 
found that the experimental data of Ref.~\onlinecite{phases} were best reproduced 
by setting the on-site impurity potential as $I_{\rm I}=-0.1$ (in eV units)~\cite{ip} 
and by using $S_{\rm I}=0$ at the impurities 
since there is evidence that Co doped in BaFe$_2$As$_2$ is non-magnetic~\cite{athena}. 
This effectively sets to zero the Hund coupling ${J_{\rm H,I}}$ 
and the spin-lattice coupling $g_{\rm I}$ at the impurity sites. 
In addition, we also reduced the localized spins to $S/4$ ($S/2$) at the NN (NNN) 
of the impurity sites with the corresponding effective decreased in $J_{\rm H}$ and $g$
at those sites. The overall chemical potential $\mu$ was adjusted so that the density of added 
impurities equals the density of added electrons, which corresponds to an ideal Co doping~\cite{phases}.  

\begin{figure}[thbp]
\begin{center}
\includegraphics[trim = 5mm 0mm 10mm 0mm,width=\linewidth,clip,angle=0]{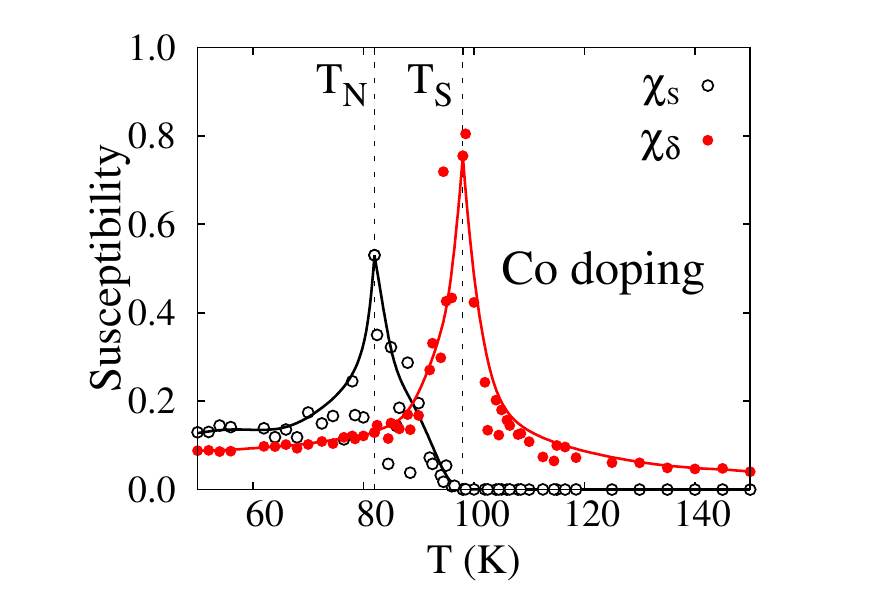}
\vskip -0.3cm
\caption{(color online) The magnetic susceptibility (open black symbols) and the lattice susceptibility (filled red symbols) vs. temperature.
The sharp peaks indicate 
the N\'eel temperature $T_N$ and the structural transition temperature $T_S$
for the case of 5\% Co-doping. The on-site disorder is $I_{\rm I}=-0.1$ 
and the off-diagonal disorder 
is defined by $S_{\rm I}=0$, $S_{\rm NN}=S/4$, and $S_{\rm NNN}=S/2$. 
The cluster used is $64\times64$.}
\label{sus}
\end{center}
\end{figure}

\begin{figure}[thbp]
\includegraphics[trim = 0mm 0mm 0mm 0mm,width=0.5\textwidth,clip,angle=0]{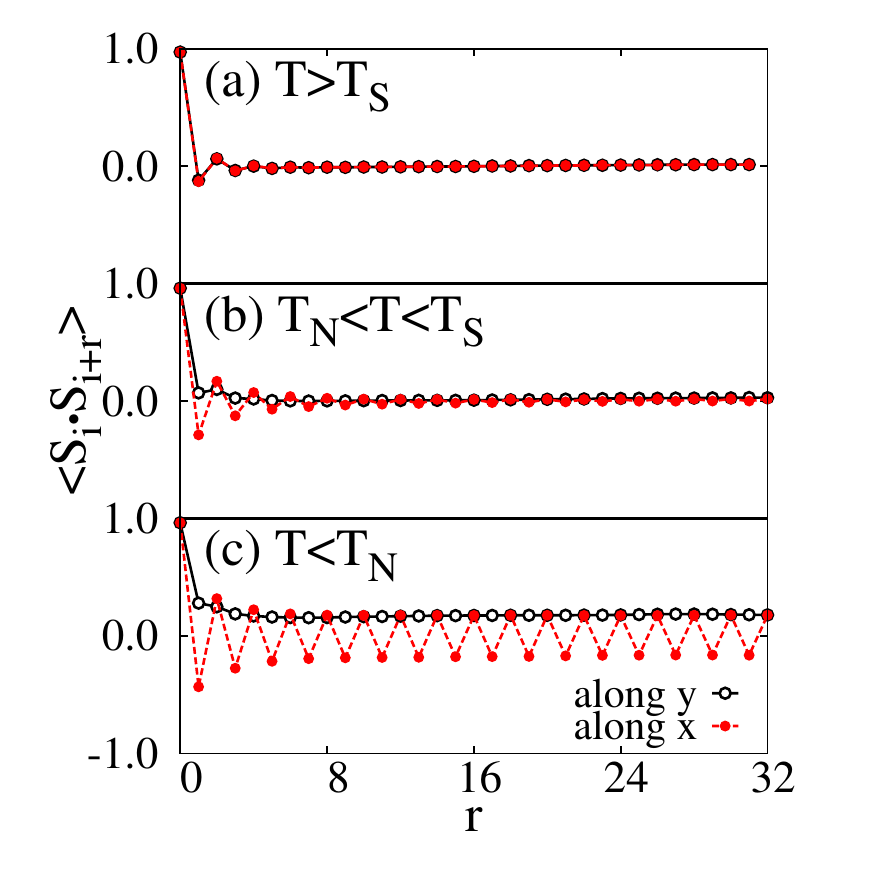}
%\includegraphics[trim = 0mm 0mm 0mm 0mm,width=9.6cm,clip,angle=0]{./Atr.pdf}
%\subfigure{\includegraphics[trim = 8mm 1mm 8mm 5mm,width=0.35\textwidth,angle=0]{./AtrT120x5.pdf}\label{spar}}
%\subfigure{\includegraphics[trim = 8mm 1mm 8mm 5mm,width=0.35\textwidth,angle=0]{./AtrT95x5.pdf}\label{snem}}\\
%\subfigure{\includegraphics[trim = 8mm 1mm 8mm 5mm,width=0.35\textwidth,angle=0]{./AtrT80x5.pdf}\label{sneel}}\\
%\subfigure[]{\includegraphics[trim = 8mm 1mm 8mm 5mm,width=0.35\textwidth,angle=0]{./phaseDCu.pdf}\label{Tx}}
%\subfigure[]{\includegraphics[trim = 8mm 1mm 8mm 5mm,width=0.35\textwidth,angle=0]{./phaseDvsn-Cu.pdf}\label{Te}}\\
\vskip -0.3cm
\caption{(color online) Real-space spin-spin correlation functions 
vs. distance on a $64\times 64$ lattice; (a) corresponds to $T=120$~K
($T>T_S$) in the paramagnetic regime, (b) to $T=95$~K ($T_N< T <T_S$) 
in the nematic state, and (c) to $T=80$~K ($T<T_N$) 
in the long-range
ordered magnetic state. The 
AFM correlations along $x$ are indicated with solid circles 
while the FM correlations along $y$ are denoted with open circles. 
The results are for 5\% Co-doping with off-diagonal disorder set by $S_{\rm I}=0$, 
$S_{\rm NN}=S/4$, and $S_{\rm NNN}=S/2$.}
\label{correl}
%\end{center}
\end{figure}

The black filled (open) circles in Fig.~\ref{xTJ} show the evolution with 
impurity doping of the structural (N\'eel) 
transition temperatures in the presence of 
the disorder caused by replacing Fe by Co at random sites. 
The magnetic dilution induced by doping causes a rapid 
reduction in $T_S$ and $T_N$, similarly as observed in 
experiments~\cite{phases}, and remarkably also opens a robust 
nematic phase for $T_N<T<T_S$ since disorder affects {\it differently} both
transition temperatures. 
In fact, the separation between $T_N$ 
and $T_S$ is very clear in the magnetic and lattice 
susceptibilities that are displayed for 5\% doping, as example, in 
Fig.~\ref{sus}. The magnetic properties of the different phases are 
also clear by monitoring the behavior 
of the real-space spin-spin correlation functions presented in Fig.~\ref{correl}. 
In panel (a) for $T=120$~K ($T>T_S$) the spin correlations effectively vanish
at distances larger than two lattice constants and there is 
no difference between the results along the $x$ and $y$ axes
directions, indicating a paramagnetic ground state. However, 
at $T=95$~K ($T_N<T<T_S$), panel (b), the correlations now 
display short-range 
AFM (FM) order along the $x$ ($y$) directions demonstrating the breakdown of the rotational invariance 
that characterizes the nematic phase, but without developing long-range order as expected. 
Lowering the temperature to $T=80$~K ($T<T_N$), panel (c), 
now the correlations have 
developed long range $(\pi,0)$ order, as expected 
in the antiferromagnetic ground state. To our knowledge,
the results in figures such as Fig.~\ref{xTJ} provide the largest separation 
between $T_S$ and $T_N$ ever reported in numerical
simulations of realistic models for iron-based superconductors.

%On-site disorder is introduced at the impurity sites by setting 
%$I_I=-0.1$ and, since Co appears non-magnetic when doped~\cite{athena} the on-site spin $S_I$ is also set to zero. In addition $S_{I,NN}$ and $S_{I_NNN}$ are set to $S/4$ 
%and $S/2$ effectively introducing off-diagonal disorder via the reduction of Heisenberg, Hund, and spin-lattice couplings in the neighborhood of the impurity sites. As it will be 
%discussed below, this is the kind of disorder that reproduces the rapid decrease of the critical temperatures and the stabilization of a nematic phase in agreement with the 
%experimental results.~\cite{phases} 

\subsection{Cu doping}

\begin{figure}[thbp]
\subfigure{\includegraphics[trim = 0mm 1mm 0mm 5mm,width=0.45\textwidth,angle=0]{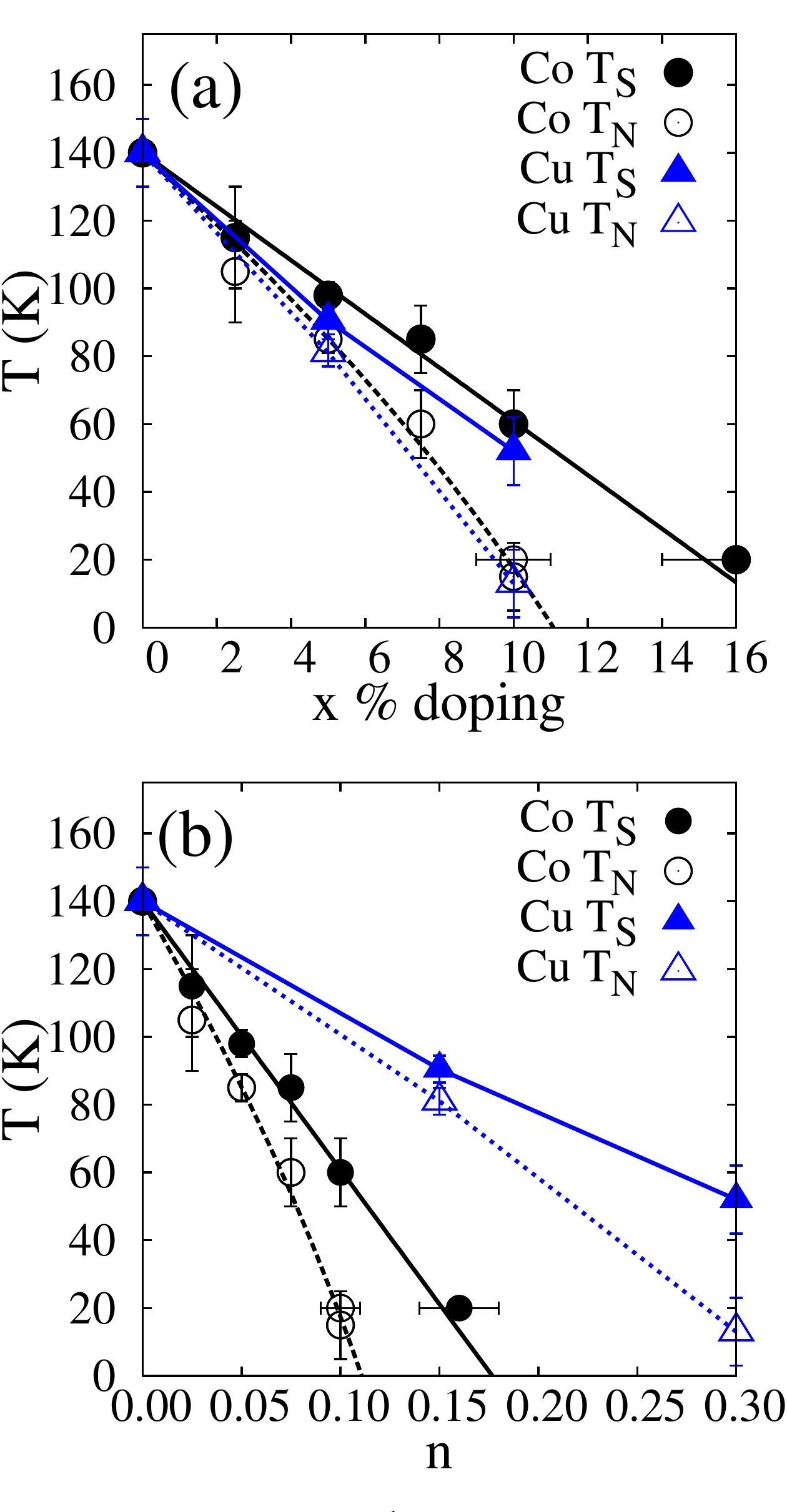}\label{Te}}
%\subfigure{\includegraphics[trim = 8mm 1mm 8mm 5mm,width=0.48\textwidth,angle=0]{./phaseDCuA.pdf}\label{Tx}}
%\subfigure{\includegraphics[trim = 8mm 1mm 8mm 5mm,width=0.48\textwidth,angle=0]{./phaseDCuB.pdf}\label{Te}}\\
%\subfigure[]{\includegraphics[trim = 8mm 1mm 8mm 5mm,width=0.35\textwidth,angle=0]{./phaseDCu.pdf}\label{Tx}}
%\subfigure[]{\includegraphics[trim = 8mm 1mm 8mm 5mm,width=0.35\textwidth,angle=0]{./phaseDvsn-Cu.pdf}\label{Te}}\\
\vskip -0.3cm
\caption{(color online) {\it Contrast of effects of Cu and Co doping.}
The N\'eel temperatures $T_N$ (dashed lines) and the structural transition temperatures $T_S$ (solid lines) for Co doping
(black open and solid circles) and for Cu doping (blue open and solid triangles) are shown. 
Results are presented first
(a) vs. the impurity density $x$ and second (b) vs. the added electronic density $n$. 
The off-diagonal disorder is set at $S_{\rm I}=0$, $S_{\rm NN}=S/4$, and $S_{\rm NNN}=S/2$. 
The cluster size is $64\times64$.}
\label{Tdop}
%\end{center}
\end{figure}

Let us consider now
the effect of doping with Cu which, nominally, introduces three electrons per doped impurity~\cite{phases}. 
For this purpose
%, we introduced disorder without changing the overall chemical potential to simulate Ru-doping and 
we increased the chemical potential at a faster rate so that the added density of
electrons is $n=3x$, instead of $n=x$ as for Co doping. 
The results are shown in Fig.~\ref{Tdop}.
When the critical temperatures for both Cu and Co doping are plotted 
as a function of the density of impurities $x$, 
in Fig.~\ref{Tdop}(a) it can be seen that the results are approximately $independent$ of 
the kind of dopant. This indicates that the critical temperatures are primarily controlled by 
the amount of quenched disorder (namely, by the number of impurity sites) 
rather than by the actual overall electronic density, at least in the range studied. 
This conclusion is in excellent agreement with the experimental phase diagrams 
shown, for example, in Fig.~26(a)
of Ref.~\cite{phases}, for the case of several transition metal oxide dopants. 
Thus, working at a fixed electronic density $n$, the values of $T_N$ and $T_S$ are smaller 
for the case of Co doping than for the case of Cu-doping, as shown in Fig.~\ref{Tdop}(b), 
because more Co than Cu impurities 
have to be added to achieve the same electronic density, underlying the fact that
Co doping introduces more disorder than Cu doping at fixed $n$. 
These results are also in good agreement 
with the experimental phase diagram in Fig.~26(b) of Ref.~\cite{phases}.

\subsection{Dependence on impurity characteristics}

\begin{figure}[thbp]
\begin{center}
\includegraphics[trim = 10mm 0mm 0mm 0mm,width=\linewidth,clip,angle=0]{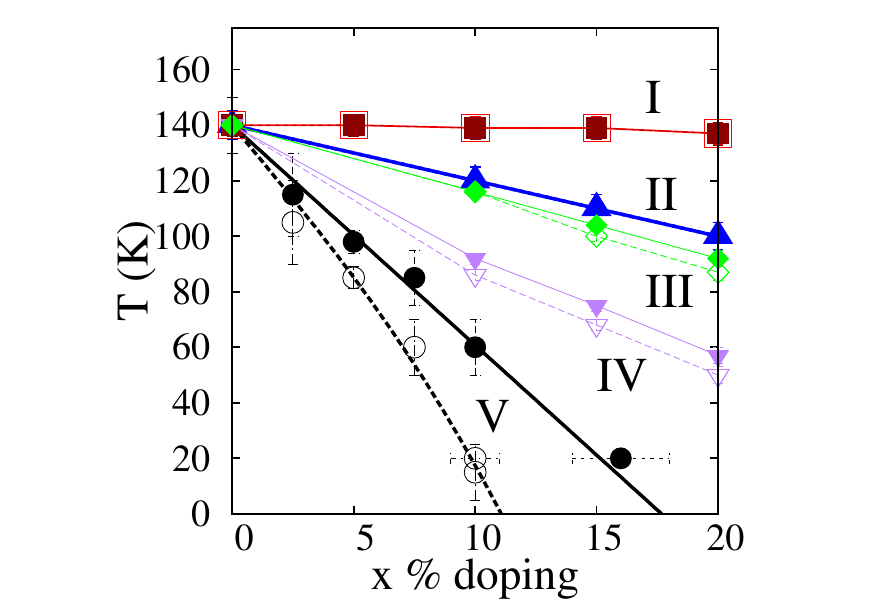}
\vskip -0.3cm
\caption{(color online) {\it Dependence of results with impurity characteristics.} 
The N\'eel transition temperature $T_N$ (dashed lines) and the structural transition temperature $T_S$ (solid lines) 
vs. the percentage of impurities $x$ for different settings of the off-diagonal 
disorder. Case I corresponds to the clean limit with no impurity sites (red squares). 
Case II has $S_{\rm I}$=$S/2$ and $S_{\rm NN}$=$S_{\rm NNN}$=$S$ untouched (blue triangles).
This case may be sufficient for Ru doping, which is magnetic. 
Case III has $S_{\rm I}$=$0$ and $S_{\rm NN}$=$S_{\rm NNN}$=$S$ untouched (green diamonds). 
Case IV has $S_{\rm I}$=$S/2$, $S_{\rm NN}$=$0.7S$, and $S_{\rm NNN}$=$0.9S$ (purple 
upside-down triangles). 
Finally, Case V has $S_{\rm I}$=$0$, $S_{\rm NN}$=$S/4$, and $S_{\rm NNN}$=$S/2$ 
(black circles). Case V appears to be the best to describe experiments for non-magnetic 
doping.
The density of doped electrons equals $x$ as in Co doping. 
In all cases the on-site disorder potential is kept fixed at $I_{\rm I}=-0.1$. The lattice 
size is $64\times64$.}
\label{xdis}
\end{center}
\end{figure}

Let us consider the dependence of the N\'eel and the structural transitions 
temperatures on the local details of the magnetic dilution caused by the disorder.
In Fig.~\ref{xdis} results for $T_N$ and $T_S$ are shown as a function of impurity 
doping with the chemical potential set to introduce one electron per dopant. 
The clean limit data (red squares, case I) 
is displayed again for the sake of comparison. The blue 
triangles (case II) are results for $I_{\rm I}=-0.1$ and $S_{\rm I}=S/2$, leaving $S_{\rm NN}$
and $S_{\rm NNN}$ untouched (i.e. equal to $S$). 
This ultra local magnetic dilution induces effective NN 
and NNN reductions in the Heisenberg couplings 
accelerating the rate of decrease of the critical 
temperatures. However, the nematic phase is still not stabilized and, thus, 
it does not reproduce the experimental behavior for the Co-doped parent compound. 
Reducing $S_{\rm I}$ to zero, as indicated by the green diamonds in the figure (case III)
and keeping $S_{\rm NN}$ and $S_{\rm NNN}$ untouched, 
slightly increases 
the rate of reduction of the critical temperatures with doping and stabilizes 
the nematic phase only after a finite amount of 
doping $x\sim 10\%$ has been added but in a very narrow range of temperature. The
conclusion of cases I, II, and III is that a very local description of the dopant
is insufficient to reproduce experiments.

We have found that in order to generate a robust 
nematic phase upon doping, extended effects of magnetic dilution 
$must$ be considered. The upside-down purple triangles (case IV) 
in Fig.~\ref{xdis} show results for $S_{\rm I}=S/2$, $S_{\rm NN}=0.7S$, 
and $S_{\rm NNN}=0.9S$. The nematic regime is still too narrow. 
But the results for  $S_{\rm I}=0$ with $S_{\rm NN}=S/4$ and $S_{\rm NNN}=S/2$ 
(black circles, case V), already shown in Fig.~\ref{xTJ}, 
indicate that increasing the strength of the extended off-diagonal disorder does 
induce a faster 
reduction of the critical temperatures and stabilizes a larger nematic region.
%of the critical temperatures and a larger region in which the nematic phase is stable (black open and filled circles in Fig.~\ref{xdis}).
Our computer simulations suggest that the range and strength of disorder, 
specifically the extended magnetic dilution, is crucial for the stabilization 
of the nematic phase when $T_N=T_S$ in the parent compound. 

We have observed that the effect of the on-site impurity 
potential $I_{\rm I}$ is weak. In principle, 
we could have kept the overall chemical potential $\mu$ fixed and control the added electronic 
density $n$ by merely 
adjusting the values of the impurity potential. However, this does not induce 
noticeable changes in the critical temperatures, due to the small overall 
modifications in the electronic density discussed before. This is not the manner in which doping seems to act in the
real electron-doped pnictides. Thus, 
we believe that working with a fixed 
value of the impurity potential and adjusting the 
electronic density with the overall chemical potential allows to study 
the effects of isotropic 
quenched disorder and varying electronic density in a more controlled and
independent way. 

Considering the negligible effect on the critical temperatures caused by pure electronic 
doping (clean limit) and, by extension, the on-site impurity potential, 
the results in Fig.~\ref{xdis} shed light on the case of 
isovalent doping in which Fe is replaced by Ru. This procedure introduces disorder but, at least nominally, 
no electronic doping. Experimental efforts have observed that in this case $T_N$ and 
$T_S$ still decrease with doping, despite no apparent changes in the Fermi surface, 
but at a slower rate than with non-isovalent doping. Moreover, the critical temperatures 
do not separate from each other, i.e., no nematic phase is stabilized~\cite{rullier}.   
% and with the phase diagram for Ru doped BaFe$_2$As$_2$ shown in Fig.~3 of Ref.~\cite{rullier}
Our results lend support to the view that the decrease of $T_N$ and $T_S$ observed with Ru-doping
is mainly due to the magnetic dilution introduced by doping rather than by more subtle effects 
on the electronic density which in turn would affect the nesting of the 
FS~\cite{hirschfeld,sawatzky}. Experiments have determined that doped Ru 
is magnetic~\cite{kim} which would translate to larger values of $S_{\rm I}$, $S_{\rm NN}$, 
and $S_{\rm NNN}$ in our model. In fact, the blue triangles (case II)
in Fig.~\ref{xdis} qualitatively capture the slower 
decrease rate and negligible separation with impurity doping for $T_N$ and $T_S$ 
experimentally observed for Ru doping~\cite{rullier}.
%The dependence of the data on the electronic density is shown in Fig.~\ref{Te} where it is clear again that the phase diagram is controlled by the magnetic 
%dilution introduced by the disorder. 

%ATTENTION!!!: do we know how <n> is affected mu_I?

\section{Properties of the Nematic Phase}\label{nematicproperties}

Having stabilized a robust nematic regime, let us study its properties.

\subsection{Neutron scattering}

Considering the importance of neutron scattering experiments in iron superconductors, 
we studied the electronic doping dependence of 
the magnetic structure factor $S({\bf k})$ obtained 
from the Fourier transform of the real-space 
spin-spin correlation functions displayed in Fig.~\ref{correl}. 
Experiments indicate that the low-temperature 
magnetic phase  below $T_S=T_N$ in the parent compound develops long range AFM (FM) order
along the long (short) lattice constant direction in the orthorhombic lattice. 
This results in a sharp peak at ${\bf k}=(\pi,0)$ (or at $(0,\pi)$ 
depending on the direction of the AFM order) 
that forms above the small spin-gap energy~\cite{xcheng}. 
More importantly for our discussion and results, 
upon electron-doping the $(\pi,0)$ neutron peak becomes broader along the direction 
transversal to the AFM order in the whole energy range~\cite{xcheng}, creating
an intriguing transverselly elongated ellipse.

\begin{figure}[thbp]
\begin{center}
\includegraphics[trim = 0mm 0mm 0mm 0mm, width=9cm,clip,angle=0]{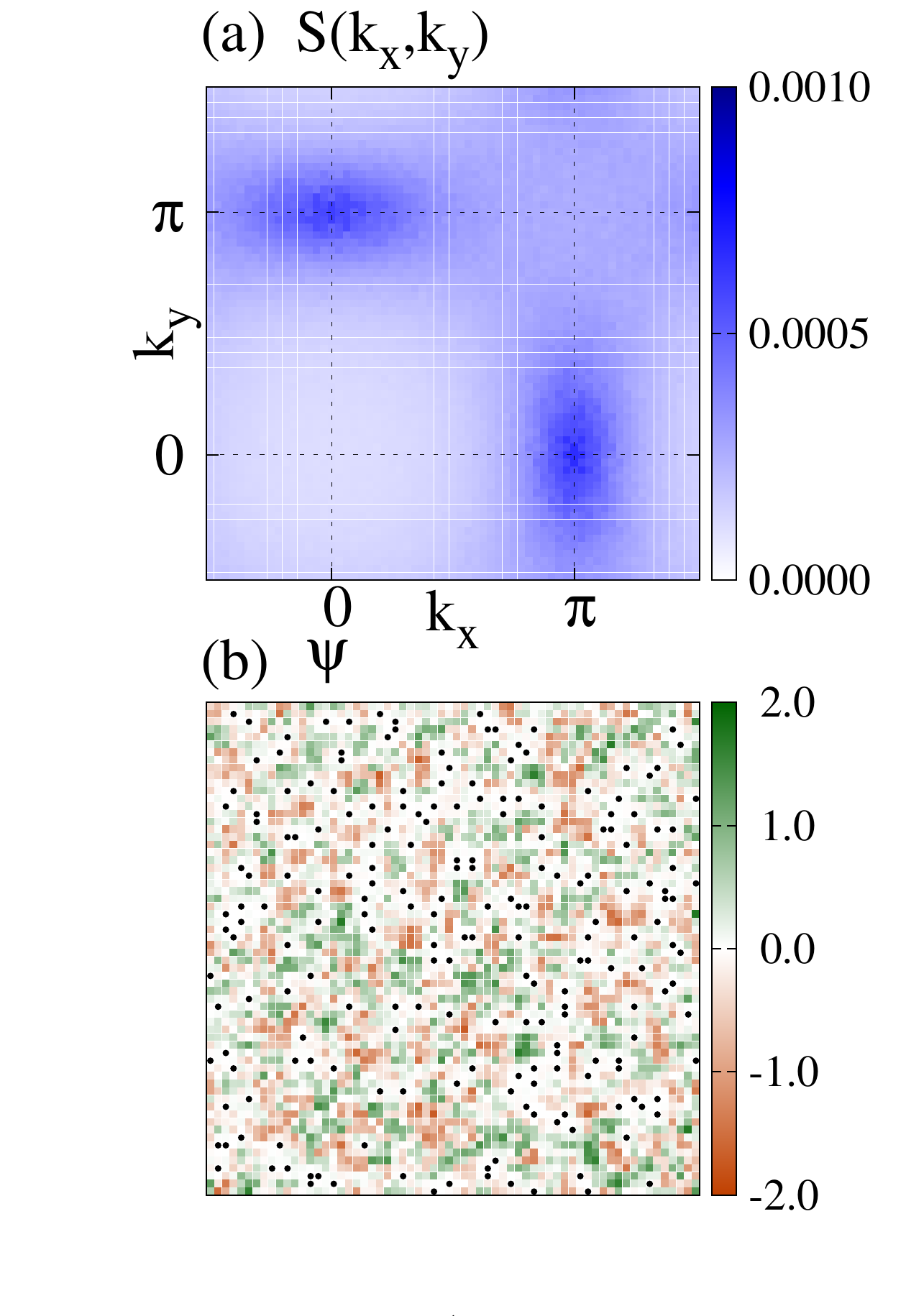}
\vskip -0.3cm
\caption{(color online) {\it Magnetic and nematic order in the paramagnetic regime.} 
The results are for 5\% Co-doping at $T=120$~K ($T>T_S$) and using a $64\times64$ lattice.
(a) The magnetic structure factor $S({\bf k})$, showing that the wavevectors 
$(\pi,0)$ and $(0,\pi)$ have similar intensity. 
(b) Monte Carlo snapshot of the spin-nematic order parameter
with approximately the same amount of 
positive (green) and negative (orange) clusters.
The impurity sites are indicated by black dots.}
\label{Tx5100}
\end{center}
\end{figure}

\begin{figure}[thbp]
\begin{center}
\includegraphics[trim = 0mm 0mm 0mm 0mm,width=9cm,clip,angle=0]{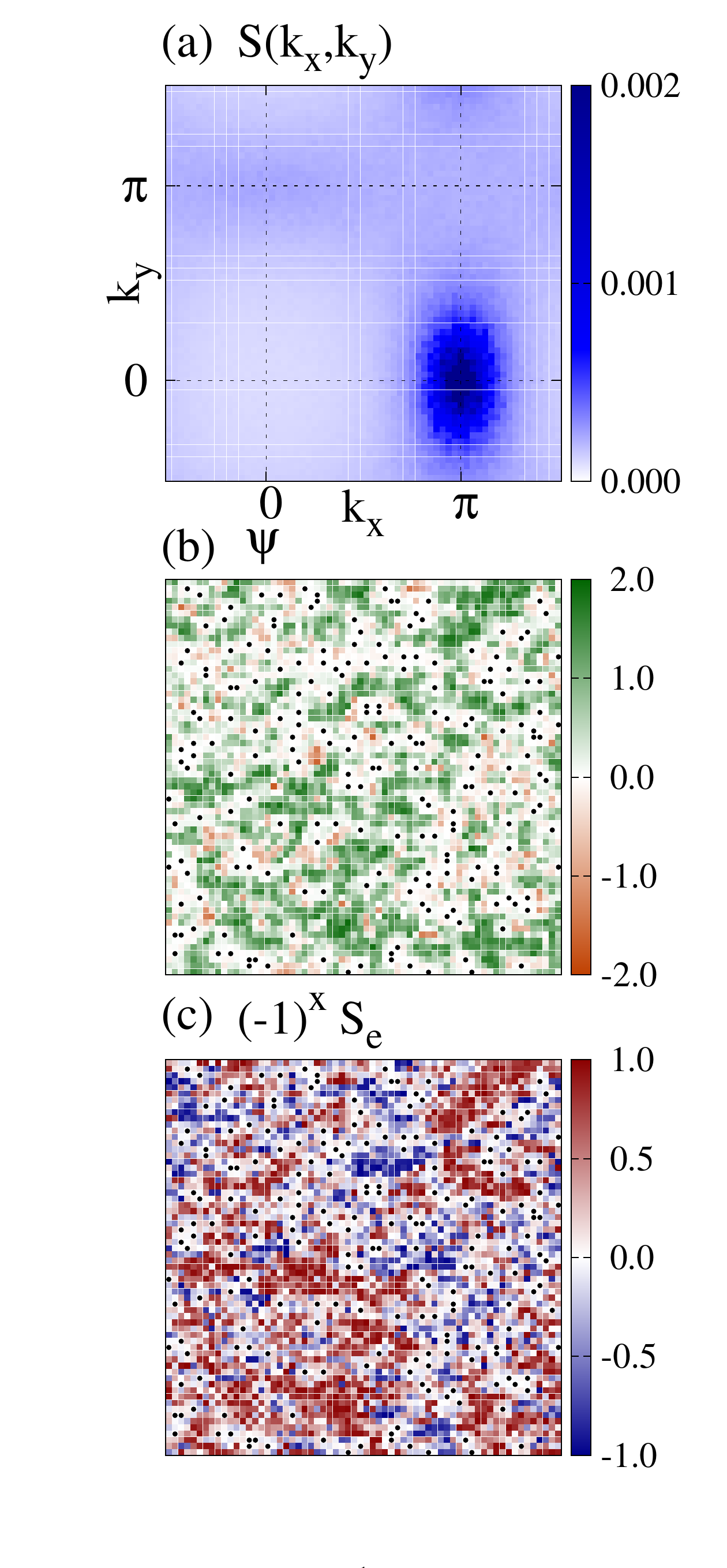}
\vskip -0.3cm
\caption{(color online) 
{\it Magnetic and nematic order in the nematic regime.} 
The results are for 5\% Co-doping at $T=95$~K ($T_N<T<T_S$) 
and using a $64\times64$ lattice.
(a) The magnetic structure factor $S({\bf k})$ is shown,
with clear dominance of the $(\pi,0)$ state. 
(b) Monte Carlo  snapshot of the spin-nematic order parameter. 
Impurity sites are indicated by black dots. 
A positive nematic order (green) dominates, but
there are still small pockets of negative order (orange).
(c) Monte Carlo snapshot displaying the on-site component along the easy axis, $S_e$, of the localized 
spin multiplied by the factor $(-1)^{{\bf i}_x}$, with ${\bf i}_x$ 
the $x$-axis component of the location of site ${\bf i}$.
Both the dominant blue and red clusters indicate regions
with $local$ $(\pi,0)$ order, but shifted 
by one lattice spacing. This shift suppresses long-range
order when averaged over the whole lattice, but short-range order remains. 
Impurity sites are denoted as black dots.}
\label{Tx590}
\end{center}
\end{figure}

The results obtained numerically for 5\% Co-doping are shown in 
Fig.~\ref{Tx5100} for $T=120$~K ($T>T_S$), i.e. 
in the paramagnetic phase. In panel (a) peaks in the spin structure factor 
$S({\bf k})$ (that represents the integral over the whole energy range of 
the neutron scattering results) with similar intensity 
at wavevectors $(\pi,0)$ and $(0,\pi)$ can be observed. Both of these peaks are 
elongated along the direction transversal to the corresponding 
spin staggered direction, in agreement with neutron scattering~\cite{xcheng}. Our  explanation for these results 
within our spin-fermion model is not associated
with Fermi Surface modifications due to electron doping, since the percentual doping
is small as already discussed, 
but instead to the development of spin-nematic clusters, 
anchored by the magnetically depleted regions that form at the 
impurity sites. A Monte Carlo snapshot of the spin-nematic order 
parameter $\Psi_{\bf i}$ on a $64\times 64$ lattice 
is shown in panel (b) of Fig.~\ref{Tx5100}. 
Since $T>T_S$, patches with $(\pi,0)$ and $(0,\pi)$ nematic order, 
indicated with green and orange in the figure, coexist in equal proportion. 
By eye inspection, we believe that the $(\pi,0)$ patches 
tend to be slightly elongated along the $x$ direction, while the $(0,\pi)$ patches 
are elongated along the $y$ direction. 
%indicating that the nematic fluctuations are stronger along the FM direction, reducing
%the size in the FM direction. 
This asymmetry could be the reason for the 
shape of the peaks in the structure factor displayed in panel (a), since
elliptical peaks can be associated to different correlation lengths
along the $x$ and $y$ axes. In Fig.~\ref{Tx5100}(a) the elliptical $(\pi,0)$ peak 
has a correlation length larger along the $x$ axis than the $y$ axis. 
 
The results corresponding to lowering the temperature 
into the nematic phase ($T=95$~K) are presented 
in Fig.~\ref{Tx590}. In this case the subtle 
effects already observed in the paramagnetic phase are magnified.
In panel (a), it is now clear that the peak at $(\pi,0)$ has
developed a much larger weight than the peak at $(0,\pi)$, as expected. 
Moreover, the elongation along the transversal direction already 
perceived in the paramagnetic state is now better developed. 
The Monte Carlo snapshot of the spin-nematic order parameter 
in panel (b) shows that the $(\pi,0)$ (green) regions prevail over the 
$(0,\pi)$ (orange) regions, indicating that the symmetry under 
lattice rotations in the nematic phase is spontaneously broken. 
In addition, now the elongated shape of the $(\pi,0)$ green clusters 
along the AFM direction is more clear to the eye. 
But despite the prevalence of $(\pi,0)$ clusters the system does not develop
long-range magnetic order (compatible with panel (b) 
of Fig.~\ref{correl}). This is because the many 
$(\pi,0)$ clusters are actually ``out of phase'' with each other. 
This is understood via the visual investigation
of Monte Carlo snapshots, as  in panel (c) of Fig.~\ref{Tx590}, where it is shown
the component of the localized spins along the easy axis, $S_e$, multiplied by a factor $(-1)^{{\bf i}_x}$ (see definition in caption; 
the location of the impurities is indicated with black dots). 
The abundant red and blue patches all indicate clusters with local $(\pi,0)$ nematic order, but shifted one with
respect to the other by one lattice spacing.
The very small regions 
with $(0,\pi)$ order, as in the orange regions of panel (b), can be
barely distinguished in panel (c) with a checkerboard red/blue structure.
% which result in 
%blue/red patches if $(-1)^y S_z$is displayed instead as shown in panel (d); in this case the $(\pi,0)$ clusters display a checkerboard structure.

%\begin{figure}[thbp]
%\subfigure[]{\includegraphics[trim = 8mm 1mm 8mm 5mm,width=0.35\textwidth,angle=0]{./T80x10B.pdf}\label{T80x10B}}
%\subfigure[]{\includegraphics[trim = 8mm 1mm 8mm 5mm,width=0.35\textwidth,angle=0]{./T50x10B.pdf}\label{T50x10B}}\\
%\caption{(Color online) (aa) The magnetic structure factor $S({\bf k})$  (ab) Snapshot on a $64\times 64$ lattice showing the impurity sites (black dots) and the on site $S_z$ component
%of the localized spin; (ac) Snapshot of the spin-nematic order parameter on  a $64\times 64$ lattice with the impurity sites indicated by black dots. The results are
%for 10\% Co-doping at $T=80K>T_S$; (b) Same as (a) but at $T_N<T=50K<T_S$.}
%\label{Sk10}
%%\end{center}
%\end{figure}

\subsection{Scanning Tunneling Microscopy}

The real space structure of the $(\pi,0)$ nematic clusters obtained numerically, 
with an elongation along the $x$ axis, 
can be contrasted with Scanning Tunneling 
Microscopy (STM) measurements. In fact, STM studies of 
Co-doped CaFe$_2$As$_2$ at 6\% doping~\cite{chuang,allan} have 
already revealed the existence of unidirectional electronic 
nanoestructures. These STM structures appear to have an average length 
of about eight lattice spacings along the AFM direction and it was argued that 
they may be possibly pinned by the Co atoms.
The picture of elongated structures along the $x$ axis is consistent 
with our results, as shown in panel (b) of Fig.~\ref{Tx590}. However, 
in our simulation the nematic structures are mainly located 
in between, rather than on top, the Co dopants. 
In our case this arises from the fact that the effect of disorder considered here 
reduces drastically the magnetic interactions at the Co or Cu dopant sites because
they are not magnetic. 

A recently discussed new perspective is that the nematic state could be
a consequence of anisotropic dopant-induced scattering rather than 
an intrinsic nematic electronic state~\cite{nakajima,ishida}, by studying the anisotropy 
in the optical spectrum~\cite{nakajima} and in the in-plane resistivity~\cite{ishida} 
varying Co doping in BaFe$_2$As$_2$. The main argument to attribute 
the observed anisotropies to extrinsic effects of Co doping is that the anisotropy 
increases with doping despite the fact that the spin order weakens  and the lattice 
orthorhombicity diminishes. Our results, by construction, were obtained with impurity
profiles that are symmetric under rotations of the lattice, so nematicity is not induced
by asymmetric Co doping characteristics. 
However, we agree with the above described experimental 
observations that quenched disorder introduced 
by the dopants is crucial for the stabilization of the nematic phase, otherwise
in the ``clean limit'' there is no difference between $T_S$ and $T_N$ as already explained.

In our simulation, the nematic phase develops because the in-plane dopants 
allowed the formation of cigar-shaped nematic domains. These domains 
have shifts in their respective AFM orders, as it can be seen in panel (c) of Fig.~\ref{Tx590}. 
%the spin in the even sites of the red and blue clusters have opposite orientations. 
%The presence of the impurities, thus, anchors the fluctuations allowing the stabilization 
%of a nematic phase above the N\'eel phase. In this case, 
%
For the 122 compounds, the dopants enhance the (weak) electronic tendency to nematicity, 
while according to our previous calculations~\cite{shuhua13} in the parent compound 
of materials in the 1111 family, such as ReFeAsO (Re= La, Nd, Sm), a small temperature 
range of nematicity can be 
provided by the coupling between the lattice and the orbital degrees of freedom. 
This view may be supported by studies of the phonon modes in the 1111 materials~\cite{dong}. 
%
%In the context of the spin-fermion model the 122 nematicity is of electronic origin, 
%but it cannot develop independently of the long-range AFM order in a robust range
%of temperatures unless the anisotropic magnetic fluctuations 
%are anchored by another degree of freedom, such as the lattice, the orbitals, or in-plane 
%impurities. 
Note also 
that atomic-resolution variable-temperature Scanning Tunnelling Spectroscopy 
experiments performed in NaFeAs, which has $T_S>T_N$, and in LiFeAs, which does not develop 
neither magnetic order nor a structural transition, 
indicate that cigar-like nematic domains develop in the nematic phase 
of NaFeAs regardless of the symmetry of the impurities observed in the 
samples~\cite{rosenthal}.

\section{Discussion and Conclusions}\label{discussion}

In this publication, the effects of electron doping in 
materials of the 122 family, such as BaFe$_2$As$_2$, 
have been investigated via numerical studies of the spin-fermion model, including 
charge, orbital, magnetic, and lattice degrees of freedom. 
These materials are electron
doped via the in-plane replacement of iron atoms by transition metal oxides, 
introducing disorder in the iron layers.
The results of our study suggest that the experimentally observed 
reduction of the magnetic and structural transition temperatures upon doping, in such
a manner that $T_N<T_S$,  
is primarily triggered by the influence of quenched disorder 
associated with the chemical substitution of magnetic Fe atoms by non-magnetic dopants 
such as Co~\cite{athena} and Cu ~\cite{singh}. More specifically, reducing 
the magnitude of the localized spins at and near
the dopants rapidly reduces 
the values of both transition critical temperatures. A ``patchy'' nematic phase is stabilized, 
which is characterized by a majority of clusters with $(\pi,0)$ order. These
patches have out-of-phase magnetic order separated by non-magnetic regions anchored by the impurities. 
While the tendency to nematicity is already a property of the purely electronic spin-fermion model, as
already discussed in previous studies~\cite{shuhua13}, it seems that for the 122 materials
this fragile tendency would not materialize into a robust nematic phase 
without the influence of disorder. Compatible with this conclusion, BaFe$_2$(As$_{1-x}$P$_x$)$_2$ (considered
among the ``cleanest'' of doped pnictides since, for example, quantum oscillations
were observed~\cite{P-doped}) has a splitting between $T_S$ and $T_N$ which is very small (if any).

Note that a mere change in chemical potential to increase the electronic doping, 
without adding quenched disorder, does $not$ stabilize 
a nematic regime and introduces at best a very small decrease 
in the transition temperatures. This
indicates that nesting effects do not play a major 
role in the opening of a robust nematic window with
doping in 122 materials. Our results also explain 
the slower decrease of the critical 
temperatures, and lack of separation between $T_N$ and $T_S$, observed 
upon Ru doping. In this case experiments have shown that 
Ru dopants in 122 materials are magnetic~\cite{kim}, contrary to
the non-magnetic nature of Co and Cu dopants. 
Thus, in our study the values of the Hund and Heisenberg 
couplings would have to be only slightly reduced at the impurity sites. 
As shown in Fig.~\ref{xdis}, this will reduce the rate of decrease, as well as the separation, 
of $T_N$ and $T_S$. The same effect may explain why $T_N=T_S$ and the decrease rate is 
slower in hole doped systems where the holes are introduced
by replacing Ba atoms reducing the effects of disorder directly in the iron layers. 

In addition, the observed clusters are elongated along the AFM direction in agreement 
with similar observations in STM experiments. Within the spin-fermion model, 
the cigar-like shape of the clusters is justified because the nearest-neighbor couplings
are AFM and, thus, fluctuations are expected to be larger along the FM (frustrated) direction 
which reduces the associated correlation length. Another consequence of this behavior is the
oval shape observed for the weight distribution 
of the magnetic structure factor around the momenta $(\pi,0)$ and $(0,\pi)$ for $T>T_N$, 
in agreement with the distribution observed in the electron-doped case in neutron scattering 
experiments. 

In summary, we report the first computational study of a realistic model for pnictides 
that reproduces the rapid drop of $T_N$ and $T_S$ with the chemical 
replacement of Fe by transition
metal elements such as Co or Cu. Since disorder affects differently $T_N$ and $T_S$, a nematic
regime is stabilized. 
The key ingredient is the introduction
of quenched disorder affecting several neighbors 
around the location of the dopant. Fermi Surface nesting
effects were found to be too small to be the main 
responsible for the fast drop of critical temperatures.
Our results are in agreement with neutron scattering and 
also with Scanning Tunneling Microscopy
that unveiled the presence of anisotropic 
nanoclusters associated with the nematic state. Considering the present results
for doped systems,  together with the previously reported  results  
for the parent compounds, we conclude  
that the spin-fermion model captures the essence of the magnetic
properties of the pnictide iron superconductors.

\section{Acknowledgments}

C.B. was supported by the National Science Foundation, under
Grant No. DMR-1404375. E.D. and A.M. were supported by the US Department of Energy, 
Office of Basic Energy Sciences, Materials Sciences and Engineering
Division.


\begin{thebibliography}{}
\bibitem{only} The spin in $H_{\rm SL}$ is only 
the localized spin for computational simplicity.
\bibitem{finsize} Smaller, and for some points, larger systems with up to $96\times 96$ sites where studied to monitor that finite size effects are very small in the $64\times 64$ clusters used to obtain the results presented here.
\end{thebibliography}
\end{document}